\documentstyle[12pt]{article}
\newcommand{\AmodG}{\overline{{\cal A}/ {\cal G}}}
\newcommand{\q}{q_0}

\begin{document}
\baselineskip=22pt plus 0.2pt minus 0.2pt
\lineskip=22pt plus 0.2pt minus 0.2pt
\font\bigbf=cmbx10 scaled\magstep3
\begin{center}
 {\bigbf Photons from quantized eletric flux representations}

\vspace*{0.35in}

\large

Madhavan Varadarajan
\vspace*{0.25in}

\normalsize

{\sl Raman Research Institute,
Bangalore 560 080, India.}
\\
madhavan@rri.ernet.in\\
\vspace{.5in}
March 2001\\
\vspace{.5in}
ABSTRACT

\end{center}
The quantum theory of $U(1)$ connections admits a diffeomorphism invariant 
representation in which the electric flux through any surface is quantized.
This representation is the analog of the representation of quantum  $SU(2)$
theory used in loop quantum gravity. We investigate the relation between this
representation, in which the basic excitations are `polymer-like', and the 
Fock representation, in which the basic excitations are wave-like photons.
We show that normalizable states in the Fock space are associated with 
`distributional' states in the quantized electric flux  representation.
This work is motivated by the question of how wave-like gravitons 
in linearised gravity arise from polymer-like states in non-perturbative 
loop quantum gravity.

\pagebreak

\setcounter{page}{1}

\section*{1. Introduction} 
The loop quantum gravity approach \cite{berniebiblio,carloreview}
is based on a Hamiltonian description of 
classical general relativity in which the basic configuration variable is a
 connection \cite{selfdual,barbero} 
and its conjugate is a triad field. The spatial 3-metric on 
a Cauchy slice is constructed from the triad and 
the connection contains information about the extrinsic curvature of the 
slice as embedded in the spacetime, thus establishing contact with the usual
Arnowitt-Deser-Misner formulation. In the quantum theory, the basic 
connection dependent operators are holonomies of the connection around loops
in the Cauchy slice \cite{carlolee,tedlee}. 

In recent years the approach has been put on a firm mathematical footing 
with the following key features. 
A (quantum) configuration space, $\AmodG$, of (generalised) $SU(2)$ 
connections modulo gauge and a canonical diffeomorphism invariant
measure, $d\mu_0$ (also called the Ashtekar-Lewandowski measure 
\cite{AL}),
on  $\AmodG$ have been constructed \cite{AI,AL,M^2,Baezproj,ALproj}. 
The space 
$L^2 ( \AmodG , d\mu_0 )$ 
provides a kinematical Hilbert space
on which the (self adjoint) $SU(2)$ holonomy operators act by multiplication.
This kinematical Hilbert space is spanned by an orthonormal set of
`spin network' states, each associated with an oriented, closed graph whose
edges are labelled by representations of $SU(2)$ \cite{RSspinnet,Baezspinnet}.

Every spin network state is an eigen state of  operators corresponding to the
area of 2-surfaces in the Cauchy slice - roughly speaking, the area of a
surface gets a contribution of $\sqrt{j(j+1)}$ units of Planck area from 
each edge labelled by spin $j$ which intersects the surface transversely.
Since the area operator is constructed from the triad field (or equivalently
the $SU(2)$ electric field), the edges of the graph may be thought of as
carrying quanta of non abelian electric flux. 
Since the edges are 1 dimensional, the intuitive picture of states in
$L^2 ( \AmodG , d\mu_0 )$ is of
 `polymer like' quantum  excitations.

As noted earlier, $L^2 (\AmodG , d\mu_0 )$ is a kinematical structure.
The dynamics of 
general relativity in its  canonical description is 
encoded in the diffeomorphism and Hamiltonian constraints.  
Physical states in the quantum theory are in the kernel of the 
corresponding quantum constraint operators. 
The representation of holonomies, functionals of the triad, as well as the 
unitary action of diffeomorphisms on $L^2(\AmodG , d\mu_0 )$ provides 
a representation of the diffeomorphism constraints \cite{ALM^2T}, and in a
key breakthrough by Thiemann, the Hamiltonian   
constraint \cite{QSD}.
It turns out that elements of the kernel of the constraints are `too 
distributional' to be normalizable states in $L^2(\AmodG , d\mu_0 )$.
Rather, they are expressible as non-normalizable, 
infinite, sums of spin network states.

Thus, a nonperturbative physical state of the quantum gravitational field is 
a distributional sum of 
kinematic states, each of which is associated with 1 dimensional,
polymer-like excitations. Given such quantum states, a key open question is:
how do classical configurations of the gravitational field arise ? 
In particular, how does flat spacetime (and small 
perturbations around it) arise from non-perturbative quantum states of the  
gravitational field?

The latter question is particularly interesting for the following reason.
Small perturbations about flat spacetime correspond to solutions of 
linearized gravity. Quantum states of linearised gravity lie in the 
familiar graviton Fock space  on which the conventional perturbative 
approaches to quantum gravity are based. Such approaches seem to fail due to
nonrenormalizability  problems. Thus, an understanding of the relation between
the quantum states of linearised gravity and states in full nonperturbative
loop quantum gravity would shed light on the reasons behind the failure of
perturbative methods. 

There are many aspects of this yet-to- be-understood relation between 
perturbative and nonperturbative states and we shall focus on only one of 
them, namely, the dramatically different nature of the basic excitations
in Fock space and in loop quantum gravity:
whereas Fock space gravitons can be thought 
of as {\em propagating 3d wave-packets},  states 
in loop quantum gravity are associated with {\em 1d polymer-like excitations}.

This aspect of the relation between perturbative and non perturbative states
in quantum gravity  can be isolated and 
studied in the simpler yet highly instructive context of source free
Maxwell theory. To this end, consider  
a $U(1)$ connection on a spatial slice diffeomorphic to $R^3$.
Starting from the Poisson bracket algebra of $U(1)$ holonomies around loops
in $R^3$, it can be seen that exact counterparts of the $SU(2)$
constructions of loop quantum gravity 
exist \cite{AI,AL,Corichi-Krasnov,AI2}. Thus, a canonical 
diffeomorphism invariant measure exists on the quantum configuration space
of generalised $U(1)$ connections modulo gauge (we shall continue to denote 
these by $d\mu_0$ and $\AmodG$ - it will be clear from the context whether
these symbols refer to $SU(2)$ or $U(1)$), holonomies act as unitary operators
by multiplication on $L^2 (\AmodG , d\mu_0 )$, and 
$L^2 (\AmodG , d\mu_0 )$ is spanned by an orthonormal basis each element 
of which is associated with a closed, oriented graph and a labelling of 
edges of the graph by representations of $U(1)$. Since such representations
are labelled by integers called charges, we refer to these states as `charge
network' states. Each such state is an eigen function of  electric flux 
operators associated with surfaces in $R^3$ and the edges of the graph 
underlying the state may be thought of as carrying quanta of electric flux.
We refer to this  diffeomorphism invariant representation, in which the 
basic excitations are, once again, 1 dimensional and `polymer-like' as 
the {\em quantized electric flux (qef) representation}.

In sharp contrast to this, is the usual Poincare invariant
quantum theory of a $U(1)$ connection on a fixed flat spacetime with flat 
spatial slices diffeomorphic to $R^3$. Here, the connection and its
conjugate electric field are represented as operator valued distributions 
on  the Fock space of photons where  the basic quantum excitations are  3d 
and wave-like.

Thus quantum  $U(1)$ theory presents an excellent arena to discuss the 
question raised earlier, namely 
{\em how can the 3d wave-like excitations of Fock space arise from
underlying 1d polymer-like excitations?} 

At first sight, the Fock and the qef representations are very different.
In the Fock representation, the connection is an operator valued distribution
which needs to be smeared in 3 dimensions to obtain a well defined operator.
Holonomies of this operator valued distribution involve smearings only 
over the 1 dimension provided by the loop, and are {\em not} well defined.
In contrast, in the qef representation, 
 holonomies {\em are} well defined operators on $L^2( \AmodG, d\mu_0 )$.
In section 2 we review our previous results \cite{me} which point to a way 
around this apparently insurmountable obstacle to relating the 
two representations.

In section 3 we present our main result, namely that states in the Fock 
representation are associated with distributional sums of charge network states
in the qef representation. This mirrors the mathematical structure of 
loop quantum gravity in the sense that, there, physical states are 
associated with distributional sums of spin network states in the kinematic 
Hilbert space. Section 4 contains a discussion of our results and some 
concluding remarks.

We shall use units in which $\hbar = c= 1$.

\section*{2. The Issue of Smearing}

Since the abelian Poisson bracket 
algebra of holonomies is the primary structure from which
the qef representation is obtained \cite{AI2,me}, we would like to
relate the Fock representation to this holonomy algebra.  As noted earlier,
the usual Fock representation  is not a representation of the holonomy algebra
because  the connection operators in this representation are too singular for
their holonomies to be defined. Instead, it is possible to obtain the usual
Fock representation as the representation of a  related algebra of 
`smeared holonomies'. By using the smeared holonomy algebra as a link 
between the algebra of holonomies  and the Fock representation we shall be able
to relate the two.

In order to define the smeared holonomies we recall the following 
definitions from \cite{lingrav,me}. $\vec x$ denotes Cartesian coordinates of
a point on $R^3$. 
$\cal{A}$ is the space of smooth $U(1)$ connections, $A_a({\vec x})$,
\footnote{
As noted in \cite{Corichi-Krasnov}, to get an object with the dimensions
of a $U(1)$ connection, we need to divide the usual magnetic potential by a
parameter, $\q$, which has dimensions of electric charge. This 
reflects in the Poisson bracket (\ref{eq:basicpb}) and the definition
(\ref{eq:afock}).}
(on the trivial $U(1)$ bundle on $R^3$) whose 
cartesian components are functions of rapid decrease at infinity.
${\cal L}_{x_0}$ is the space of unparametrized, oriented, piecewise 
analytic loops  
on $R^3$ with basepoint $\vec{x_0}$. 
Composition of a loop $\alpha$ with a loop $\beta$ is denoted by 
$\alpha\circ \beta$.
Given a loop $\alpha \in {\cal L}_{x_0}$,
the holonomy of $A_a(x)$ around $\alpha$ is 
$H_\alpha (A):= \exp (i \oint_{\alpha} A_a dx^a)$.
The holonomy can equivalently be defined as 
\begin{equation}
H_{\alpha}(A)  =  \exp i\int_{R^3} X^a_{\gamma}({\vec x}) 
A_a({\vec x}) d^3x, 
\label{eq:ha}
\end{equation}
with 
\begin{equation}
X^a_{\gamma}({\vec x}) :=  
\oint_{\gamma} ds \delta^3({\vec {\gamma}} (s), {\vec x}){\dot{\gamma}}^a,
\label{eq:x}                                
\end{equation}
where $s$ is a parametrization of the loop $\gamma$, 
$s\in [0,2\pi]$.
$X^a_{\gamma}({\vec x})$ 
is called the form factor  of $\gamma$. 
The Gaussian smeared form factor \cite{lingrav} is defined as
\begin{equation}
 X^a_{\gamma_{(r)}}({\vec x}) := 
\int_{R^3} d^3y f_r({\vec y}-{\vec x})
            X^a_{\gamma}({\vec y}) 
  =  \oint_{\gamma} ds f_r({\vec \gamma (s)}- {\vec x}){\dot{\gamma}}^a
\label{xr}
\end{equation}
where
\begin{equation}
f_r({\vec x}) = {1\over (2\pi )^{3\over 2}r^3}e^{-x^2\over 2r^2} 
\;\;\;x:=|{\vec x}| 
\end{equation}
approximates the Dirac delta function for small $r$.

Then, the smeared holonomy is defined as 
\begin{equation}
H_{\gamma_{(r)}}(A)  = \exp i\int_{R^3} 
X^a_{\gamma_{(r)}}({\vec x}) A_a({\vec x}) d^3x .
\label{eq:hr}
\end{equation}

As shown in \cite{me}, the Fock representation  
is a representation 
of the Poisson bracket algebra generated by the smeared holonomies, 
$H_{\gamma_{(r)}}(A)$ and the electric field $E^a({\vec x})$:
\begin{eqnarray}
\{ H_{\gamma_{(r)}},H_{\alpha_{(r)}}\} &=
       & \{ E^a({\vec x}), E^b({\vec y})\} =0,  \nonumber\\
\{ H_{\gamma_{(r)}}, E^a({\vec x})\} &= & {i\over \q}
                                          X^a_{\gamma_{(r)}}({\vec x})
                                          H_{\gamma_{(r)}} .
\label{eq:pbr}
\end{eqnarray}
These Poisson brackets are generated from the elementary Poisson brackets
\begin{equation}
\{ A_{a}({\vec x}), E^b ({\vec y})\} = {1\over \q}
                   \delta_a^b \delta ({\vec x},{\vec y}).
\label{eq:basicpb}
\end{equation}
Here, {\em $\q$ is a parameter with the units of electric charge} 
\cite{Corichi-Krasnov}.  
In the Fock representation, 
the smeared holonomies, ${\hat H}_{\gamma_{(r)}}$,
are unitary operators and the electric field, 
${\hat E}^a({\vec x})$, is an operator valued distribution.

How is the above algebra involving smeared holonomies and the electric field
 related to the holonomy algebra? 
To this end we 
define  the classical Gaussian smeared electric field $E^a_r({\vec x})$ by 
\begin{equation}
E^a_r({\vec x}) := 
\int d^3y f_r({\vec y}-{\vec x})
            E^a({\vec y}) .
\label{eq:defer}
\end{equation}
The Poisson bracket algebra
generated by the (unsmeared) holonomies and the Gaussian smeared 
electric field is 
\begin{eqnarray}
\{ H_{\gamma},H_{\alpha}\} &=
       & \{ E^a_r({\vec x}), E^b_r({\vec y})\} =0,  
\nonumber
\\
\{ H_{\gamma}, E^a_r({\vec x})\} &= & {i\over \q}
               X^a_{\gamma_{(r)}}({\vec x})
                                          H_{\gamma} .
\label{eq:pb}
\end{eqnarray}

Then, as is hinted by (\ref{eq:pbr}) and 
(\ref{eq:pb}) and proved in detail in 
\cite{me}, the abstract algebraic structures underlying the Poisson
bracket algebras generated by
$(H_{\gamma_{(r)}}(A) , E^a({\vec x}))$ and 
$(H_{\gamma}(A) , E^a_r({\vec x}))$ are {\em identical}!
In other words, 
$E^a({\vec x})$ and $E^a_r({\vec x})$, and $H_{\gamma_{(r)}}(A)$ and 
$H_{\gamma}(A)$ may be identified with the same abstract objects as 
far as the algebraic structure of the two Poisson bracket algebras is 
concerned. Therefore, any representation of the Poisson bracket algebra 
generated by $(H_{\gamma_{(r)}}(A) , E^a({\vec x}))$ defines a 
representation of the Poisson bracket algebra generated by 
$(H_{\gamma}(A) , E^a_r({\vec x}))$. 
In particular, the Fock representation of the 
$(H_{\gamma_{(r)}}(A) , E^a({\vec x}))$ algebra defines a 
representation of the $(H_{\gamma}(A) , E^a_r({\vec x}))$ algebra.
We shall call this representation, the `$r$-Fock representation' of 
the $(H_{\gamma}(A) , E^a_r({\vec x}))$ Poisson bracket algebra.
Thus, in the  $r$-Fock representation the holonomies,
${\hat H}_{\gamma}$ {\em are} well defined operators!

We can see this explicitly as follows. The Fock representation, or for that 
matter, any representation with a cyclic `vacuum' state, 
can be reconstructed from the vacuum expectation  values of the algebra
of operators. Thus, we may specify the $r$-Fock representation via the 
vacuum expectation values of the Fock representation as follows.
The Fock representation is a cyclic representation generated from 
the Fock vacuum, $|0>$ with vacuum expectation values
\footnote{Our conventions for the Fock space representation are 
displayed in section 3.2}
\begin{eqnarray}
<0|{\hat H}_{\gamma_{(r)}}|0> &=&
\exp (-\int {d^3k \over 4\q^2 k} 
            | X^a_{\gamma_{(r)}}({\vec k})|^2), 
\label{eq:fockvevh} \\
<0|{\hat H}_{\alpha_{(r)}}{\hat E}^a({\vec x}){\hat H}_{\beta_{(r)}}|0> &=&
{X^a_{\beta_{(r)}}({\vec x})-X^a_{\alpha_{(r)}}({\vec x})\over 2\q}
\exp (-\int {d^3k \over 4\q^2 k} 
            | X^a_{{\alpha\circ \beta}_{(r)}}({\vec k})|^2),
\label{eq:fockvev}
\end{eqnarray}
where $X^a_{\gamma_{(r)}}({\vec k})$ denotes the Fourier transform of 
$X^a_{\gamma_{(r)}}({\vec x})$.
This defines the $r$-Fock 
representation as the cyclic representation generated from the $r$-Fock 
`vacuum' $|0_r>$ with vacuum expectation values
\begin{eqnarray}
<0_r|{\hat H}_{\gamma}|0_r>&=&
 \exp -(\int {d^3k \over 4\q^2 k} 
            | X^a_{\gamma_{(r)}}({\vec k})|^2). 
\label{eq:rfockvevh}
\\
<0_r|{\hat H}_{\alpha}{\hat E}^a_r({\vec x}){\hat H}_{\beta}|0_r> &=&
{(X^a_{\beta_{(r)}}({\vec x})-X^a_{\alpha_{(r)}}({\vec x}))\over 2 \q}
\exp (-\int {d^3k \over 4\q^2 k} 
            | X^a_{{\alpha\circ \beta}_{(r)}}({\vec k})|^2).
\label{eq:rfockvev}
\end{eqnarray}
The fact that the 
holonomies are well defined operators in the $r$-Fock representation allows us 
to related the $r$-Fock representation and the qef representation in a fairly 
direct manner, as we show in the next section.

To summarise: Although it is not possible to relate the qef representation 
with the usual Fock representation directly, it {\em is} 
possible to relate the 
qef representation with the $r$-Fock representation, since both provide 
representations of the Poisson bracket algebra generated by 
$(H_{\gamma}(A) , E^a_r({\vec x}))$.

The only remaining question is of the relation between the Fock and the 
$r$-Fock representations. The mathematical relation between the two, in terms
of representations of 
two different realizations of the same algebraic structure, is clear and is 
true for {\em any} $r>0$. So the only question is of the {\em physical}
relation between $r$-Fock and Fock representations. We argue below that 
for certain measurements which are of physical relevance in the context of 
our motivations from quantum gravity, the two representations are 
physically indistinguishable for sufficiently small $r$.

Since we shall phrase our argument in terms of measurements of Fourier modes,
we first discuss the behaviour of Fourier modes under Gaussian smearing.
Given any function $h({\vec x})$, its Fourier transform is 
\begin{equation}
h({\vec k})= {1\over (2\pi)^{3\over 2}}\int_{R^3}d^3x h({\vec x})
e^{-i{\vec k}\cdot {\vec x}}
\label{eq:ft}
\end{equation}
and its Gaussian smeared version is
\begin{equation}
h_r({\vec x})=\int_{R^3} d^3y f_r({\vec y}-{\vec x})
            h({\vec y}) .
\label{eq:r}
\end{equation} 
It follows that
\begin{equation}
h_r ({\vec k}) = e^{-k^2r^2 \over 2}h({\vec k}) .
\label{eq:rk}
\end{equation}
In particular we have 
\begin{equation}
X^a_{\gamma_{(r)}}({\vec k}) = e^{-k^2r^2 \over 2}X^a_{\gamma}({\vec k})
\label{eq:xrk}
\end{equation}
and 
\begin{equation}
E^a_r ({\vec k}) = e^{-k^2r^2 \over 2}E^a({\vec k}) .
\label{eq:erk}
\end{equation}
From (\ref{eq:xrk}) and (\ref{eq:hr}) it follows, in obvious notation, that
\begin{equation}
H_{\gamma_{(r)}}(A({\vec k}))= H_{\gamma}( e^{-k^2r^2 \over 2}A({\vec k})).
\label{eq:ark}
\end{equation}

Consider measurements of quantities at or above a length scale $L$. 
More precisely, let the measurements be of Fourier modes, 
$E^a({\vec k}), A_a({\vec k})$ of the electric and connection fields (in, say, 
the Coulomb gauge) for 
$k\leq {1\over L}$. Further, let the accuracy of the measurement process 
be characterised by the positive  number $\delta$. If 
$\Delta E^a({\vec k}), \Delta A_a({\vec k})$ are the accuracies to which
$E^a({\vec k}), A_a({\vec k})$ are measured, then 
$\delta$ is defined through $\Delta E^a({\vec k})=  E^a({\vec k})\delta$
and $\Delta A_a({\vec k})=  A_a({\vec k})\delta$. Then if
$r$ is chosen small enough that the condition
$1-e^{-{r^2\over 2L^2}} <\delta$ holds, it follows that the 
measurements cannot distinguish between the modes $E^a({\vec k})$
and  $e^{-{k^2r^2\over 2}}E^a({\vec k})$,
and $A_a({\vec k})$ and $e^{-{k^2r^2\over 2}}A_a({\vec k})$.
Thus, the measurements cannot distinguish between unsmeared fields and their
Gaussian smeared  counterparts for sufficiently small $r$. Since the 
 the primary operators in the Fock and $r$-Fock representations are related 
by Gaussian smearing (see equations (\ref{eq:erk}) and (\ref{eq:ark})),
it is straightforward to see that for any state in the Fock space 
there exists a state in the $r$-Fock space such that the above type of
measurement can never distinguish between the two.
\footnote{In the 
language and notation of \cite{me}, the state in the $r$-Fock space is the 
image of the state in the Fock space via the map $I_r$.}

We loosely interpret this statement to mean that, given measurements at 
some length scale performed to some finite accuracy, there is always 
a sufficiently small $r$ such that the $r$-Fock 
representation is experimentally
indistinguishable from the usual Fock representation.
Although for Maxwell theory the introduction of a length scale seems arbitrary,
in the context of linearised gravity it is necessary to restrict attention 
to length scales much larger than the Planck scale or else the 
linearised approximation will not be physically valid.

Hence, for the remainder of the paper we may alter our original question
to : {\em how can the 3d wave-like excitations of $r$-Fock space arise from
underlying 1d polymer-like excitations?} 

\section*{3. $r$-Fock states as distributions in the qef representation}
In this section the $r$-Fock space representation is derived from the 
qef representation by identifying the $r$-Fock vacuum as a distributional
state in the qef representation. We show that the $r$-Fock vacuum can be 
written as a formal non-normalizable sum of charge network states and 
that it resides in the algebraic dual to the space of finite 
linear combinations of charge network states.

In section 3.1 we briefly review the properties of charge network states
as well as the definition of the algebraic dual representation.
In section 3.2.1 we encode the Poincare invariance of the Fock vacuum 
in a relation between smeared holonomy and electric field operators.
In section 3.2.2 we show how this relation implies the identification 
of the $r$-Fock vacuum with a distributional sum of charge network states.

\subsection*{3.1. Charge network states}
Straightforward repetitions of constructions for the $SU(2)$ case of 
loop quantum gravity \cite{RSspinnet,ALproj,Baezproj,Baezspinnet}
lead to the following results for $U(1)$ charge networks.

\noindent {\bf (1)} The charge network states constitute  an (uncountable)
orthonormal spanning set in 
$L^2( \AmodG , d\mu_0 )$. Each such state is labelled by a closed, oriented
graph whose edges carry non-trivial
representations of $U(1)$. Representations of 
$U(1)$ are labelled by integers called `charges' \cite{Corichi-Krasnov},
hence the name `charge network states'. $U(1)$ gauge invariance implies that 
the sum of charges at each vertex vanishes. We denote the normalized state
labelled by the graph `$\gamma$', with $N$ edges carrying the charges
$(p_1..p_N)$  as $|\gamma , \{ p\}>$. Orthonormality implies 
\begin{equation}
<\alpha,  \{ q\}|\gamma , \{ p\}> = 
\delta_{(\alpha,  \{ q\}),(\gamma , \{ p\})} 
\label{eq:onormal}
\end{equation}
i.e. the inner product vanishes unless $\alpha =\gamma$ and 
$q_i= p_i, \;i=1..N$.

\noindent  {\bf (2)} The electric field operator acts as
\begin{equation}
{\hat E}^a( {\vec x} ) |\gamma , \{ p\}>
= {X^a_{\gamma ,\{ p\}}({\vec x})\over \q} |\gamma , \{ p\}>.
\label{eq:ecn}
\end{equation}
Here 
\begin{equation}
X^a_{\gamma ,\{ p\}}({\vec x}) :=  
\sum_{i=1}^N 
p_i\int_{e_i} ds_i \delta^3({\vec e}_i (s_i), {\vec x}){\dot{e_i}}^a .
\label{eq:xe}                                
\end{equation}
In this equation $e_i$ denotes the $i$th edge of $\gamma$ and is   
parametrized by the parameter $s_i$ so that 
${\vec e}_i (s_i)$ is the coordinate 
of the point on the edge $e_i$ at
parameter value $s_i$. 

The electric flux operator associated with a surface $S$ with surface normal
$n_a$ acts as
\begin{equation}
\int_S {\hat E}^a n_a d^2s |\gamma , \{ p\}>
= {\sum_{i_S}^{}p_{i_S} \kappa_{i_S}\over \q}|\gamma , \{ p\}> .
\label{efcn}
\end{equation}
Here the index $i_S$ ranges only over the edges of $\gamma$ which intersect
the surface $S$. $\kappa_{i_S} =1$ iff ${\dot{e_{i_S}}}^an_a >0$,
$\kappa_{i_S} =-1$ iff ${\dot{e_{i_S}}}^an_a <0$ and
$\kappa_{i_S} =0$ iff ${\dot{e_{i_S}}}^an_a =0$. This action may be 
derived 
rigorously via a regularization along the lines of \cite{ALarea}.

From (\ref{efcn}), every charge network state $|\gamma , \{ p\}>$ is an
eigen state of the electric flux operator and the edges of $\gamma$
may be pictured as carrying quanta of electric flux in multiples of 
${\q}^{-1}$.
Although physical intuition for (and, indeed, the naming of) the qef 
representation arises from this property of the electric flux, 
the electric flux operator itself 
will not play a role in the considerations of 
this work. Instead, it is the Gaussian smeared electric field operator
which we will play a key role and we now exhibit its action on charge network
states.

In the notation of (\ref{xr}), it follows that
(\ref{eq:ecn}) implies 
\begin{equation}
{\hat E}^a_r( {\vec x} ) |\gamma , \{ p\}>
= {X^a_{{\gamma ,\{ p\}}_{(r)}}({\vec x})\over \q} |\gamma , \{ p\}>.
\label{eq:ercn}
\end{equation}

In addition to {\bf (1)} and {\bf (2)}, the abelian nature of $U(1)$
implies the following.

\noindent {\bf (3)} Every charge network state $ |\gamma , \{ p\}>$
can be obtained from the ``vacuum'' state $|\Omega >$ (i.e. the state
$\Omega (A) = 1$, $\Omega \in L^2( \AmodG , d\mu_0 )$) via the action of 
the holonomy operator around a suitably defined loop, $\beta $, so that
\begin{equation}
|\gamma , \{ p\}> = {\hat H}_{\beta} |\Omega > .
\label{eq:graphloop}
\end{equation}
Here $\beta = \beta_1^{p_1}\circ\beta_2^{p_2}...\circ\beta_N^{p_N}$
and 
$\beta_i^{p_i}$ denotes the loop obtained by traversing $p_i$ times
round $\beta_i$. $\beta_i,\; i= 1..N$ are defined by the construction 
(3.2) of \cite{AL} as 
\begin{equation}
\beta_i = Q (v_i^+) \circ e_i \circ Q (v_i^-) 
\end{equation}
where $v^{\pm}_i$ are the vertices of $\gamma$ which constitute the 
beginning and end points of $e_i$ and $Q(v)$ is a path from the 
base point ${\vec x}_0$ to the point $v$ such that $Q(v)$ intesects
$\gamma$ at most at a finite number of isolated points. It can be verified 
that 
\begin{equation}
X^a_{\gamma ,\{ p\}}({\vec x})= X^a_{\beta}({\vec x})
\end{equation}
and that (\ref{eq:graphloop}) holds. For this reason we denote 
${\hat H}_\beta$ by ${\hat H}_{\gamma ,\{ p\}}$.

Conversely, it can be checked that for any loop $\beta\in {\cal L}_{x_0}$,
$X^a_{\beta}({\vec x})=X^a_{\gamma ,\{ p\}}({\vec x})$, where the closed,
oriented graph $\gamma$ is the union of the edges which comprise $\beta$ with 
the orientations of the edges in $\gamma$ chosen arbitrarily 
and the labelling $\{ q\}$, given such a choice of orientation is as follows.
Let the number of times an edge $e_i$ is traversed in $\beta$, in the same
direction as its orientation in $\gamma$,  be  $q_{1i}$.
Let the number of times the  edge $e_i$ is traversed in $\beta$, in the 
opposite
direction to its orientation in $\gamma$,  be  $q_{2i}$. Then $e_i$ is
labelled by $q_i = q_{1i}-q_{2i}$. Henceforth, we shall use the labelling of
holonomies by their associated charge networks (i.e. $H_{\gamma,\{ p\}}$)
  interchangabley 
with their labelling by loops (i.e.$ H_\beta$). Thus, 
if there is 
no charge labelling in the subscript to $H$, the label is to be understood
as a loop else as an associated charge network.

\noindent {\bf (4)} Consider the holonomy operator associated with 
charge network label $(\alpha, \{ q\} )$.
\footnote{Strictly speaking the discussion should and can be framed in terms 
of holonomically equivalent labels (i.e. $(\alpha, \{ p \} )$ is equivalent to
$(\beta, \{ q \} )$ iff $X^a_{\alpha , \{ p \}}=X^a_{\beta , \{ q \}}$).
We gloss over this subtlety in the interest of pedagogy.}
 Then
${\hat H}_{\alpha , \{q\}}$ maps 
$|\gamma , \{ p\}>$ to a new charge network state based on the 
graph $\gamma \cup \alpha$ consisting of the union of the 
sets of  edges belonging
to $\gamma$ and  $\alpha$.  
\footnote{It is assumed that edges of $\alpha , \gamma$ overlap only if they 
are identical and that intersections of $\alpha , \gamma$ occur only at 
vertices of $\alpha , \gamma$ . This entails no loss of 
generality, since we can always find graphs which are 
holonomically equivalent to $\alpha , \gamma$ and for which the 
assumption holds.}
The edges
of $\gamma \cup \alpha$ are oriented and labelled with charges as follows.
Edges which are not shared by $\gamma$ and $\alpha$ retain their orientations
and charge labels. Any shared edge labelled by the charge $p$ in $\gamma$
retains its orientation from $\gamma$ and has charge $p+q$ if it has the 
same orientation in $\alpha$ and charge $p-q$ if it has opposite orientation 
in $\alpha$. We denote this new state by $|\gamma \cup \alpha ,\{p\cup q\}>$.
Thus 
\begin{equation}
{\hat H}_{\alpha , \{ q\} } |\gamma , \{ p\}>
= |\gamma \cup \alpha ,\{p\cup q\}> .
\label{eq:hcn}
\end{equation}
It can be checked that (\ref{eq:onormal}),(\ref{eq:ercn}) and 
(\ref{eq:hcn}) define a representation of the Poisson bracket algebra 
of equation(\ref{eq:pb}) such that ${\hat H}_{\gamma}$ is  unitary  and 
${\hat E}^a_r({\vec x})$ is self adjoint.

Next, we review the  construction of the `dual'  representation on the 
space of algebraic duals. Let ${\cal D} \subset L^2( \AmodG , d\mu_0 )$ be the 
(dense) set of finite linear combinations of charge network states.
Let ${\cal D}^*$ be the space of algebraic duals to ${\cal D}$ i.e.
every $\Phi \in {\cal D}^*$ is  a complex linear map on ${\cal D}$.
Define the action of ${\hat E}^a_r({\vec x}), {\hat H}_{\alpha}$ on 
${\cal D}^*$ through 
\begin{eqnarray}
({\hat E}^a_r({\vec x})\Phi ) (|\gamma , \{ p\}>) & := &
\Phi ( {\hat E}^a_r({\vec x})^{\dagger}|\gamma , \{ p\}>)
= \Phi ( {\hat E}^a_r({\vec x})|\gamma , \{ p\}>) ,
\label{eq:dualer} \\
({\hat H}_{\alpha}\Phi ) (|\gamma , \{ p\}>) & := &
\Phi ({\hat H}_{\alpha}^{\dagger}|\gamma , \{ p\}>)
= \Phi ({\hat H}_{\alpha^{-1}}|\gamma , \{ p\}>).
\label{eq:dualh}
\end{eqnarray}
Equations (\ref{eq:dualer}) and (\ref{eq:dualh}) provide an 
(anti-)representation of the Poisson bracket algebra (\ref{eq:pb}).
Note that ${\cal D}^*$ merely provides a linear representation space
for the dual representation - it does  not inherit  any natural 
inner product from ${\cal D}$. 

In section 3.2 we shall see that the natural arena to discuss the relation 
between the $r$-Fock representation and the qef representation
is the space of algebraic duals,${\cal D}^*$.  

\subsection*{3.2 The condition of Poincare invariance}
Even in linear  quantum  field theory there is no analog of the Stone- von 
Neuman uniqueness theorem for quantum mechanics on a vector space. Hence 
there are infinitely many inequivalent representations of the Poisson bracket 
algebra of smeared holonomies (see equation (\ref{eq:pbr})). The Fock space 
representation is singled out by the additional requirement of 
Poincare invariance, which in turn, is encoded in the specific choice of
complex structure (i.e. the positive-negative frequency decomposition)
for the Fock representation. This choice is equivalent to the requirement that 
the Fock vacuum be a zero eigenstate of the Fock space annihilation operators.
In section 3.2.1 we shall express this requirement as a relation between
the action of the smeared holonomy and electric field operators on the 
Fock vacuum (see equation (\ref{eq:pif})). 

The image of this relation 
in the $r$-Fock representation (see equation (\ref{eq:pirf}))
may be thought of as a condition which picks out the $r$-Fock vacuum.
In other words, the $r$-Fock vacuum may be thought of as a solution to the 
condition (\ref{eq:pirf}). Since this condition relates the action
of the holonomy and the smeared electric field operators, it is well defined
in {\em any} representation of the Poisson bracket algebra of equation
(\ref{eq:pb}). In section 3.2.2 we impose the condition (\ref{eq:pirf}) in the 
(dual-)qef representation and show that it has a {\em unique} (upto a
multiplicative constant) solution!
Since this condition is deduced from the requirement of Poincare invariance,
we may interpret the solution of (\ref{eq:pirf}) as the $r$-Fock vacuum 
expressed as a state in the (dual-)qef representation.

\subsection*{3.2.1 Poincare invariance in terms of smeared holonomies}

Our conventions for the Fock space representation are as follows.
The expansions of the field operators in Coulomb gauge are 
\begin{eqnarray}
{\hat A}_a ({\vec x})& =& 
{1\over \q {(2\pi )}^{3\over 2}}
\int {d^3k\over \sqrt{k}} (e^{i{\vec k}\cdot {\vec x}} 
{{\hat a}_a({\vec k})\over\sqrt{2}}
             + \;\; {\rm hermitian}\;{\rm conjugate}), 
\label{eq:afock}
\\
 {\hat E}_a ({\vec x})& =& 
{1\over {(2\pi )}^{3\over 2}}
\int d^3k\sqrt{k} (-ie^{i{\vec k}\cdot {\vec x}} 
{{\hat a}_a({\vec k})\over\sqrt{2}}
             + \;\; {\rm hermitian}\;{\rm conjugate}).
\label{eq:efock}\\
\Rightarrow  {\hat H}_{{\alpha, \{ q\}}_{(r)}}& =& 
 \exp \left({i\over \q}
      \int {d^3k\over \sqrt{2k}}X^a_{{\alpha, \{ q\}}_{(r)}}({\vec k})
({\hat a}_a({\vec k}) +{\hat a}^{\dagger}_a({\vec k}))\right)              
\label{eq:hfock}
\end{eqnarray}
The commutation relation between the annihilation and creation operators is 
\begin{equation}
[{\hat a}_a({\vec k}),{\hat a}_b^{\dagger}({\vec l})]= 
(\delta_{ab}-{k_ak_b\over k^2})
                                           \delta ({\vec k}, {\vec l}). 
\label{eq:comm}
\end{equation}
Using these equations, the vacuum expectation value of the smeared holonomy
operator (also called, in the language of \cite{me}, `the Fock Positive Linear 
Functional' or the Fock PLF and denoted by 
$\Gamma_F$) evaluates to 
\footnote{Note that the expression for $\Gamma_F$ in \cite{me} is incorrect.
The correct expression is (\ref{eq:fockplf}) above and differs from the 
expression in \cite{me} by a factor in its exponent. Also, in
\cite{me} the parameter $\q^{-1}$ was written as $e$, but factors of 
$e$ appeared in \cite{me} often in the wrong places. We have corrected the
erroneous expressions of \cite{me} in this work. With appropriate
corrections regarding these factors of $\q$, all the results of
\cite{me} continue to hold.}
\begin{equation}
\Gamma_F ([\alpha , \{ q \} ]):=<0 | H_{{\alpha, \{ q\}}_{(r)}}|0> 
= \exp (-{1\over 4 \q^2} \int {d^3k\over k}|X^a_{{\alpha, \{ q\}}_{(r)}}
                                        ({\vec k})|^2)
\label{eq:fockplf}
\end{equation}
With equation (\ref{eq:comm}), 
equations (\ref{eq:efock}) and (\ref{eq:hfock}) provide a representation 
of the Poisson bracket algebra (\ref{eq:pbr}). As discussed 
above the condition of Poincare 
invariance implies  
\begin{equation}
{\hat a}_a({\vec k}) |0> =0 .
\label{eq:a0}
\end{equation}
This, in turn, implies the following relation between the 
smeared holonomy and the electric field operators:
\begin{equation}
(\Gamma_F ([\alpha , \{ q \} ]) )^2 
e^{-\int {d^3x\over \q} ({1\over \sqrt{-(\partial^c\partial_c)}}
         X^a_{{\alpha, \{ q\}}_{(r)}}({\vec x})){\hat E}_a ({\vec x})} |0>    
=  {\hat H}^{\dagger}_{{\alpha, \{ q\}}_{(r)}}|0> .
\label{eq:pif}
\end{equation}
This equation holds for {\em every} closed oriented graph
 $\alpha$ and encodes the 
condition of Poincare invariance in terms of elements of the 
algebra of smeared holonomies and the electric field.

\subsection*{3.2.2 Poincare invariance in terms of holonomies}
The image of condition (\ref{eq:pif}) in the $r$-Fock representation is 
\begin{equation}
(\Gamma_F ([\alpha, \{ q \}]) )^2 
e^{-\int {d^3x\over \q} ({1\over \sqrt{-(\partial^c\partial_c)}}
         X^a_{{\alpha, \{ q\}}_{(r)}}({\vec x}))
{\hat E}_{ra} ({\vec x})} |0_r>    
=  {\hat H}^{\dagger}_{\alpha, \{ q\}}|0_r> .
\label{eq:pirf}
\end{equation}

We impose this condition in the qef representation 
on the space ${\cal D}^*$ of algebraic duals defined in section 3.1.
Thus, the following equation is to be solved for some $\Phi_0 \in {\cal D}^*$:
\begin{equation}
(\Gamma_F ([\alpha]) )^2 e^{-\int {d^3x\over \q} 
({1\over \sqrt{-(\partial^c\partial_c)}}
         X^a_{{\alpha, \{ q\}}_{(r)}}
({\vec x})){\hat E}_{ra} ({\vec x})} \Phi_0    
=  {\hat H}^{\dagger}_{\alpha, \{ q\}}\Phi_0 .
\label{eq:piqef}
\end{equation}
Any element of ${\cal D}^*$ can be written as a formal sum over all charge 
network states as follows. If 
$\Phi_0 (|\gamma , \{ p\}>) =c_{\gamma,\{ p\}} $ 
where  $c_{\gamma,\{ p\}}$ is a
complex number, it follows that $\Phi_0$ can be written as
\begin{equation}
\Phi_0 = \sum_{\gamma, \{ p\}}c_{\gamma,\{ p\}}<\gamma , \{p\}| .
\label{eq:cgammap}
\end{equation}
Substitution of this in  (\ref{eq:piqef}) and projection of the resulting 
equation onto the ket $|\beta,  \{ t\}>$ yields
\begin{eqnarray}
(\Gamma_F ([\alpha, \{ q \} ]) )^2 
\sum_{\gamma, \{ p\}}c_{\gamma,\{ p\}}<\gamma , \{p\}|
e^{-\int {d^3x\over \q} ({1\over \sqrt{-(\partial^c\partial_c)}}
         X^a_{{\alpha, \{ q\}}_{(r)}}({\vec x})){\hat E}_{ra} ({\vec x})} 
  |\beta,  \{ t\}> & \nonumber \\
= \sum_{\gamma, \{ p\}}c_{\gamma,\{ p\}}<\gamma , \{p\}| 
                                        {\hat H}_{\alpha, \{ q\}} 
|\beta,  \{ t\}>& .
\end{eqnarray}
Orthonormality of the charge network states, together with 
(\ref{eq:ercn}) and (\ref{eq:hcn}), implies that 
\begin{equation}
(\Gamma_F ([\alpha, \{ q \} ]) )^2 
e^{-\int {d^3x\over \q^2} ({1\over\sqrt{-(\partial^c\partial_c)}}
         X^a_{{\alpha, \{ q\}}_{(r)}}({\vec x})) 
X_{a{\beta, \{ t\}}_{(r)}}({\vec x})} c_{\beta,\{ t\}}
=c_{\beta \cup \alpha ,\{ t\cup q \}}
\label{eq:csol}
\end{equation}

This equation can now be solved for the coefficients $c_{\gamma,\{ p\}}$;
of course, since the equation is linear and homogeneous the solution will be
ambiguous by an overall constant. We fix this ambiguity by setting 
the coefficient labelled by the trivial graph $\gamma = 0$,  
${\vec 0}(s)= {\vec x}_0$,  to be unity.
Then setting $\beta = 0$ 
 in (\ref{eq:csol}) yields 
\begin{equation} 
c_{\alpha, \{ q \}} = (\Gamma_F ([\alpha, \{ q \} ]) )^2
\end{equation}
for every charge network $|\alpha, \{ q \}>$.
It may be verified that, miraculously, 
this also provides a solution to (\ref{eq:csol})!
Thus 
\begin{equation}
\Phi_0 = \sum_{\gamma, \{ p\}}(\Gamma_F ([\gamma, \{ p \} ]) )^2
                                         <\gamma , \{p\}| .
\label{eq:cunique}
\end{equation}
is the {\em unique}  (upto an overall constant) 
solution to the condition (\ref{eq:piqef})!
We identify $\Phi_0$ as the state corresponding to the $r$-Fock vacuum. 

As
shown in \cite{me} the action of the smeared holonomy operators on the 
Fock vacuum generates a dense subset of the Fock space. It follows that the 
action of the holonomy operators on the $r$-Fock vacuum generates 
a dense set of the $r$-Fock space. Therefore, we can use the dual 
representation of the holonomy operator (see equation (\ref{eq:dualh}))
on ${\cal D}^*$ to generate the corresponding set of states from
$\Phi_0$. Call this set ${\cal L}^*$. Thus any element of 
${\cal L}^*$ is of the form 
$\sum_{I=1}^N a_I {\hat H}_{\gamma^I, \{ p^I \}} \Phi_0$ for
some complex $a_I$ and  $N$ finite.

As noted earlier, ${\cal D}^*$ (and hence ${\cal L}^*$) is not equipped 
with an inner product. Therefore an inner product on ${\cal L}^*$ must be 
chosen which implements the classical `reality conditions', namely
$H_{\gamma, \{ p \}}^*= H_{\gamma, \{ -p \}}$ and 
$E_r({\vec x})^* =E_r({\vec x}) $,
\footnote{ Rather than using these reality conditions directly for the
operator ${\hat E}_r({\vec x})$, it is simpler to use   them to induce 
adjointness relations on 
the operators 
$e^{-\int {d^3x\over \q} ({1\over \sqrt{-(\partial^c\partial_c)}}
         X^a_{{\alpha, \{ q\}}_{(r)}}({\vec x})){\hat E}_a ({\vec x})}$.}
 on the corresponding quantum operators.
It can be verified that the following inner product (naturally 
extendible to all  of ${\cal L}^*$) implements the reality conditions:
\begin{equation}
( {\hat H}_{\alpha, \{ p \}}\Phi_0, {\hat H}_{\beta, \{ q \}}\Phi_0)
 = \exp (-{1\over 4\q^2} 
\int {d^3k\over k}|X^a_{{\alpha\cup\beta, \{ -p\cup q\}}_{(r)}}
                                        ({\vec k})|^2) .
\label{eq:fockip}
\end{equation}
It follows that the  
Cauchy completion of ${\cal L}^*$ with respect to this  inner product 
results in a Hilbert space which can be identified as $r$-Fock space and 
that the 
representation given by equations (\ref{eq:dualer}) and (\ref{eq:dualh})
is exactly the $r$-Fock representation.

\section*{4. Concluding remarks}

In this work we have related the diffeomorphism invariant, non-seperable
quantized electric flux representation for quantum $U(1)$ theory 
to its  standard Poincare invariant 
Fock space representation. This relation is based  on the fact that the $U(1)$ 
holonomies play an important role in the construction of the qef 
representation.

Since the holonomy operators are well defined in the qef representation but
{\em not} in the Fock representation, we first constructed a 1 parameter 
family of  representations in which the holonomy operators {\em are} well
defined and which are physically indistinguishable from the standard Fock
representation. More precisely, the new representations are labelled by a
positive parameter, $r$, with dimensions of length. For finite accuracy 
measurements at distance scales much larger than $r$, these `$r$-Fock
representations' are indistinguishable from the standard Fock representation.

Next, we related the $r$-Fock representation (for any fixed $r$) to the qef 
representation by identifying the $r$-Fock vacuum as a (distributional) state
in the (dual) qef representation. This identification was achieved by solving,
in the qef representation, equation
(\ref{eq:piqef}) inspired by Poincare invariance, which 
enforced the condition that the 
annihilation operator of the $r$-Fock representation 
kill its vacuum state. 
The qef representation is built on the property of diffeomorphism  
invariance
and `knows' nothing about Poincare invariance and hence
we find it truly remarkable that the condition (\ref{eq:piqef}) which arises
from Poincare invariance of the Fock vacuum can be solved
essentially uniquely in the (dual) qef representation.
Once the $r$- Fock vacuum was identified as a state in the dual qef 
representation, we constructed states corresponding to a dense set in
$r$-Fock space by the (dual) action of the holonomy operators on the 
$r$-Fock vacuum. Finally, the inner product was obtained on this set of states
by requiring that the classical reality conditions be implemented as 
adjointness conditions on the corresponding quantum operators. Thus, 
in the qualitative language of the introduction section, we may say that 
non-normalizable infinite superpositions  
of 1 dimensional, polymer like excitations
conspire, in collusion with the inner product (\ref{eq:fockip}), to
acquire the character of 3d wavelike excitations in Fock space.

From  the point of view of quantum gravity, we think that we have unravelled 
an important set of structures which will help relate the Fock space of
gravitons of linearised gravity to appropriate semiclassical states in loop
quantum gravity. In loop quantum gravity also, it is the {\em dual}
representation to the kinematic spin network representation which serves as
the `home' for physical, dynamically relevant  quantum states of the 
gravitational field. Although there is no consensus on the exact physical 
states of the theory, it is still true that the {\em structure} of the 
physical states is that of distributions on the finite span of spin network
states and that quantum operators act on these distributions via dual action.
Further, the issue of the correct inner product on the space of physical 
states is still open and this inner product may have very different properties
from the kinematical one in the context of solutions to the 
Hamiltonian constraint.
Since the $r$-Fock representation for $U(1)$ theory has been obtained, in this
work, as the dual representation on distributions to the finite span of charge
network states with the `physical' inner product (\ref{eq:fockip}) unrelated to
the `kinematic' inner product (\ref{eq:onormal}), we feel that our results
will play an important role in relating gravitons to physical states in 
loop quantum gravity.
In the loop quantum gravity case there are other complications such as the 
`linearization' of the non abelian gauge group to 3 copies of $U(1)$ 
\cite{lingrav}, as well as the identification of a state corresponding to
flat spacetime. These issues are currently under investigation.

Apart from potential applications to quantum gravity, it would be of interest
to understand the  $r$-Fock representations in their own right. In this regard
we raised the question in \cite{me} as to whether the $r$-Fock representation 
could be realised as an $L^2( \AmodG , d\mu_{F(r)} )$ representation
for some measure
$d\mu_{F(r)}$ on $\AmodG$.  As we show in the appendix, the answer to this
question is in the affirmative and it would be of interest to understand 
the properties of this new $r$-Fock measure. Whether these new representations
have applications outside of loop quantum gravity remains to be seen.

\section*{Acknowledgments}

\noindent 
I gratefully acknowledge helpful discussions of
this material with Abhay  Ashtekar. I am indebted to Marcus van Bers
for sharing his insights with me.

\section*{Appendix}
\subsection*{A1 Existence of the $r-$Fock measure on $\AmodG$}
In what follows we shall freely use results and notation from \cite{me} as
well as from previous sections. 
\footnote{We shall use the corrected expressions of \cite{me} - see 
footnote 6 in this regard.}
We shall be brief - the interested reader may
work out the details.
The $r$- Fock measure exists on $\AmodG$ iff the $r$-Fock positive linear 
functional, 
\begin{equation}
\Gamma_{F(r)} (\sum_{I=1}^{N}a_I[ \alpha_I ]) = \sum_{I=1}^N a_I
\exp (-{1\over 4\q^2} \int {d^3k\over k}|X^a_{{\alpha_I}_{(r)}}
                                        ({\vec k})|^2)
\end{equation}
is continuous with respect to the $C^*$ norm, 
$||\sum_{I=1}^N a_I [\alpha_I ]|| := 
\sup_{A\in {\cal A}} |\sum_{I=1}^N a_I H_{\alpha_I}(A)|$. From \cite{me}
we have that 
\begin{equation}
||\sum_{i=1}^N a_I [\alpha_I ]||= 
\sup_{A\in {\cal A}} |\sum_{I=1}^N a_I H_{{\alpha_I}_{(r)}}(A)| .
\label{eq:zero}
\end{equation}
Also, from (\ref{eq:fockvevh}) we have that 
\begin{equation}
\Gamma_{F(r)} (\sum_{I=1}^{N}a_I[ \alpha_I ]) = 
<0|\sum_{I=1}^N a_I{\hat H}_{\alpha_{I(r)}}|0>.
\label{eq:one}
\end{equation}
Using the standard $L^2 ({\cal S}^*, d\mu_G)$ \cite{me,glimmjaffe}
representation of Fock space where ${\cal S}^*$ is an appropriate space
of tempered distributions  and $d\mu_G$ is the standard, unit volume,
Gaussian measure, we have
\begin{equation}
<0|\sum_{I=1}^Na_I{\hat H}_{\alpha_{I(r)}}|0> = 
\int_{A_a\in {\cal S}^*} d\mu_G(A) \sum_{I=1}^Na_I H_{\alpha_{I(r)}}(A).
\end{equation}
\begin{eqnarray}
\Rightarrow
|<0|\sum_{I=1}^Na_I{\hat H}_{\alpha_{I(r)}}|0>| &\leq & 
           \sup_{A\in {\cal S}^*} |\sum_{I=1}^N a_I H_{{\alpha_I}_{(r)}}(A)|
                     \int_{A_a\in {\cal S}^*} d\mu_G(A)\nonumber\\    
                    &= & 
\sup_{A\in {\cal S}^*} |\sum_{I=1}^N a_I H_{{\alpha_I}_{(r)}}(A)|. 
\label{eq:two}
\end{eqnarray}
Since $X^a_{\alpha_{(r)}}({\vec x})$ is in Schwartz space, it follows that 
every $A_a\in {\cal S}^*$ defines a homeomorphism, $h$, from 
${\cal HG}_r$ to $U(1)$ (the element of $U(1)$ corresponding to a loop 
$\alpha$ is just $\exp i\int_{R^3} 
X^a_{\alpha_{(r)}}({\vec x}) A_a({\vec x}) d^3x $). 
It follows from the considerations of \cite{me} (see especially 
equation (A15) of \cite{me}) that 
\begin{equation}
\sup_{A\in {\cal S}^*} |\sum_{I=1}^N a_I H_{{\alpha_I}_{(r)}}(A)|
\leq \sup_{A\in {\cal A}} |\sum_{I=1}^N a_I H_{{\alpha_I}_{(r)}}(A)|.
\label{eq:three}
\end{equation}
It follows from (\ref{eq:zero}), (\ref{eq:one}), (\ref{eq:two})
and (\ref{eq:three}) that 
\begin{equation}
|\Gamma_{F(r)} (\sum_{I=1}^{N}a_I[ \alpha_I ])| 
\leq ||\sum_{i=1}^N a_I [\alpha_I ]|| .
\label{eq:four}
\end{equation}
This implies that $\Gamma_{F(r)}$ is continuous with respect to $||\;\;||$
and hence that an $r$-Fock measure, $d\mu_{F(r)}$ exists on $\AmodG$.

\end{document}